\documentclass[10pt,twocolumn,aps,prl,superscriptaddress,showpacs,byrevtex,floatfix]{revtex4-2}
\usepackage{epsf,graphicx,amssymb}
\usepackage[usenames]{color}
\usepackage{natbib}
\usepackage{ulem}
\usepackage{amsmath,amssymb}
\usepackage{color}
\usepackage{empheq}
\usepackage{url}
\usepackage{placeins}

\begin{document}
\title{Internal Waves Control Bulk Flow in Silos}

\author{David Luce}
\email{david.luce@univ-lorraine.fr}
\affiliation{LEMTA, Université de Lorraine, CNRS, 2, Avenue de la Forêt de Haye, Vandœuvre-lès-Nancy, 54504, France} 
\affiliation{GRASP, Institute of Physics B5a, Université de Liège, 4000 Liège, Belgium}
\author{Adrien Gans}
\affiliation{LEMTA, Université de Lorraine, CNRS, 2, Avenue de la Forêt de Haye, Vandœuvre-lès-Nancy, 54504, France}
\author{Nicolas Vandewalle} 
\affiliation{GRASP, Institute of Physics B5a, Université de Liège, 4000 Liège, Belgium}
\author{Sébastien Kiesgen De Richter}
\email{sebastien.kiesgen@univ-lorraine.fr}
\affiliation{LEMTA, Université de Lorraine, CNRS, 2, Avenue de la Forêt de Haye, Vandœuvre-lès-Nancy, 54504, France}
\affiliation{Institut Universitaire de France (IUF)}

\begin{abstract}{
We experimentally measure paticle acceleration within the bulk during the discharge of a granular silo. We highlight the existence of a deceleration wave emerging at the outlet level near the dead zone and propagates toward the top of the medium. The wave emission frequency is extracted from spatiotemporal diagrams of the Eulerian instantaneous acceleration profiles. Surprisingly, we find that this frequency decreases with the cohesion of the medium and is independent of the outlet size.}
\end{abstract}

\maketitle

Granular media frequently show complex behavior in a variety of geophysical phenomena and industrial applications, such as additive manufacturing or food processing \cite{andreotti2013granular, barletta2007solid, cannavacciuolo2009arch}. Among the various applications of granular materials, silos are one of the most commonly used types of equipment in the industry for storage and handling \cite{Hagen1852,beverloo1961flow, litwiniszyn1964application, mullins1974experimental, nedderman1979kinematic, medina1998velocity, mankoc2007flow, bazant2003theory}. Amidst the complexity of hopper flows, which have been extensively studied and characterized through the identification of semi-empirical laws governing various aspects, such as the influence of aperture size on flow rate \cite{Hagen1852,beverloo1961flow, mankoc2007flow} and the morphology of velocity profiles both at the outlet \cite{janda2012flow, BEN14} and in the bulk \cite{mullins1974experimental, bazant2003theory}, the occurrence of "silo quaking" emerges as a recurrent challenge in the realm of silo discharge dynamics. Gravity-driven flows within silos typicaly manifest as a cyclic or pulsating motion, influenced by fluctuations in material density during flow and varying degrees of internal friction mobilization and boundary wall friction \cite{sperl2006experiments, xu2022modified, perge2012evolution}. This pulsating flow phenomenon becomes notably pronounced at lower flow rates, exhibiting periods ranging from mere seconds to extended durations, potentially extending to minutes \cite{zuriguel2005jamming, zuriguel2014clogging, zuriguel2019velocity}. The repercussions of such pulsating loads vary, ranging from minor nuisances like shock wave transmission through the ground to disturbances in neighboring areas, to the more severe structural fatigue failures induced when the flow pulses excite the natural frequencies of the silo and its supporting structure \cite{carson2001silo, wilde2008silo}. A related issue to the silo quaking problem is the phenomenon known as "silo music" and "silo honking," as reported by Tejchman and Gudehus \cite{tejchman1993silo}. Silo honking, specifically, is observed in tall aluminum silos storing plastic powders, where higher-frequency components of flow pulsations generate loud, periodic "fog-horn" type sounds, causing notable disturbances \cite{muite2004silo}.

Through an experiment conducted in a 2D silo, we demonstrate the existence of internal waves and characterize them. Specifically, we examine the frequency of these waves and show that it is independent of the aperture size, contrary to the scaling laws describing the flow rate. Additionally, we find that the frequency of these waves is strongly influenced by the cohesion of the medium. We interpret these results in light of the formation of free-fall arch in the lower region of the silo, at the interface between the flowing and the dead zones.

\begin{figure}[!h]
    \centering
    \vskip -0.15cm
    \includegraphics[width=0.49\textwidth, trim=0 0cm 0 0, clip]{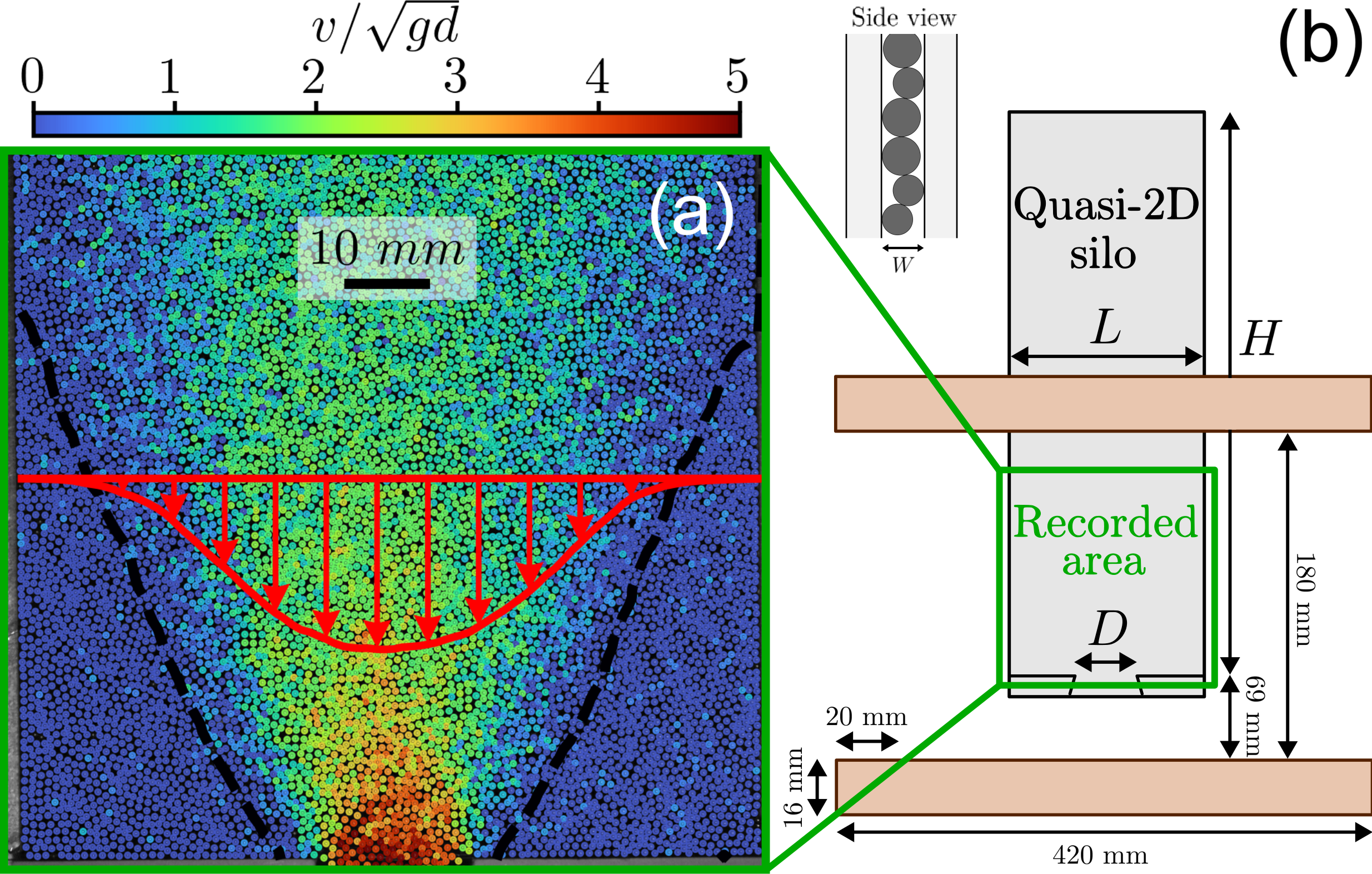}
    \vskip -0.1cm
    \caption{(a) Flow near the silo outlet with the corresponding velocity profile (shown in red). Black dashed lines indicate the boundaries of the dead zones, defined by $v/\sqrt{gd} < 0.05$. (b) Schematic of the experimental setup.} 
    \label{Fig:sketch_silo2D}
\end{figure}

The experimental setup is a quasi-2D silo (shown in Fig.~\ref{Fig:sketch_silo2D} (b)) identical to that used by Pascot et al. \cite{pascot2020influence,pascot2022discharge} in their study of the influence of vibrations on silo discharge. It consists of two conductor polycarbonate plates which prevent electrostatic effects. The dimensions of the cavity are $H = 300$~mm (height), $L = 100$~mm (width) and $W = 1.25$~mm (depth) with an adjustable aperture size $D$ set here from $6$~mm to $24$~mm. The narrow depth of the silo allows for a monolayer of an amorphous 2D stack of spherical particles. When the silo is discharging, grains fall directly into a collecting container.

\vskip -0.15cm

The flow is recorded by a high-speed camera with a frame rate of $2000$~Hz and exposure time of $498$~µs. Pictures (1024×1024 pixels) show a $10 \times 10 $~cm$^2$ area at the bottom of the quasi-2D silo (see Fig.~\ref{Fig:sketch_silo2D} (a)). Grains are uniformly illuminated and individually identified using a particle tracking algorithm. The granular media used is composed of bidisperse steel beads of diameter $1.0$~mm and $1.2$~mm (mean diameter $d = 1.1$~mm $\pm$ $0.1$~mm). Coils in a Helmholtz configuration are used to apply a uniform magnetic field across the silo opening. When placed in a magnetic field, beads acquire magnetic dipolar moments and behave like small magnets. They attract each other in the vertical direction and repel each other in the horizontal direction. The induced cohesion is controlled by adjusting the current $I$ passing through the coils (up to $4$~A here).

To quantify interparticle cohesion, we define a magnetic Bond number as the ratio of the magnetic force to the gravitational weight of a single grain, following the standard approach used in magnetic cohesion–dominated systems \cite{vandewalle2009flow, thorens2021discharge, forsyth2001effect}. By measuring the angle of repose of the dead zones formed after discharge, which ranges from 28° to 55°, we estimate Bond numbers between 0 and 51.1, consistent with previously reported results \cite{forsyth2001effect}.\\

In this work, we operate in a regime where the outlet size is sufficiently large to prevent complete flow arrest due to the formation of stable arches. Unlike previous studies \cite{zuriguel2014clogging, TO01, gella2017linking, Nicolas18} focusing on the jamming events occurring at the silo exit, our analysis examines flow dynamics within the bulk of the granular material. In this region, we observe velocity profiles exhibiting the characteristic shape shown in Fig. \ref{Fig:sketch_silo2D} (b), with a central, channelized flowing zone bordered by two quasi-static dead zones.

\begin{figure}[!h]
    \centering
    \vskip -0.1cm
    \includegraphics[width=0.48\textwidth]{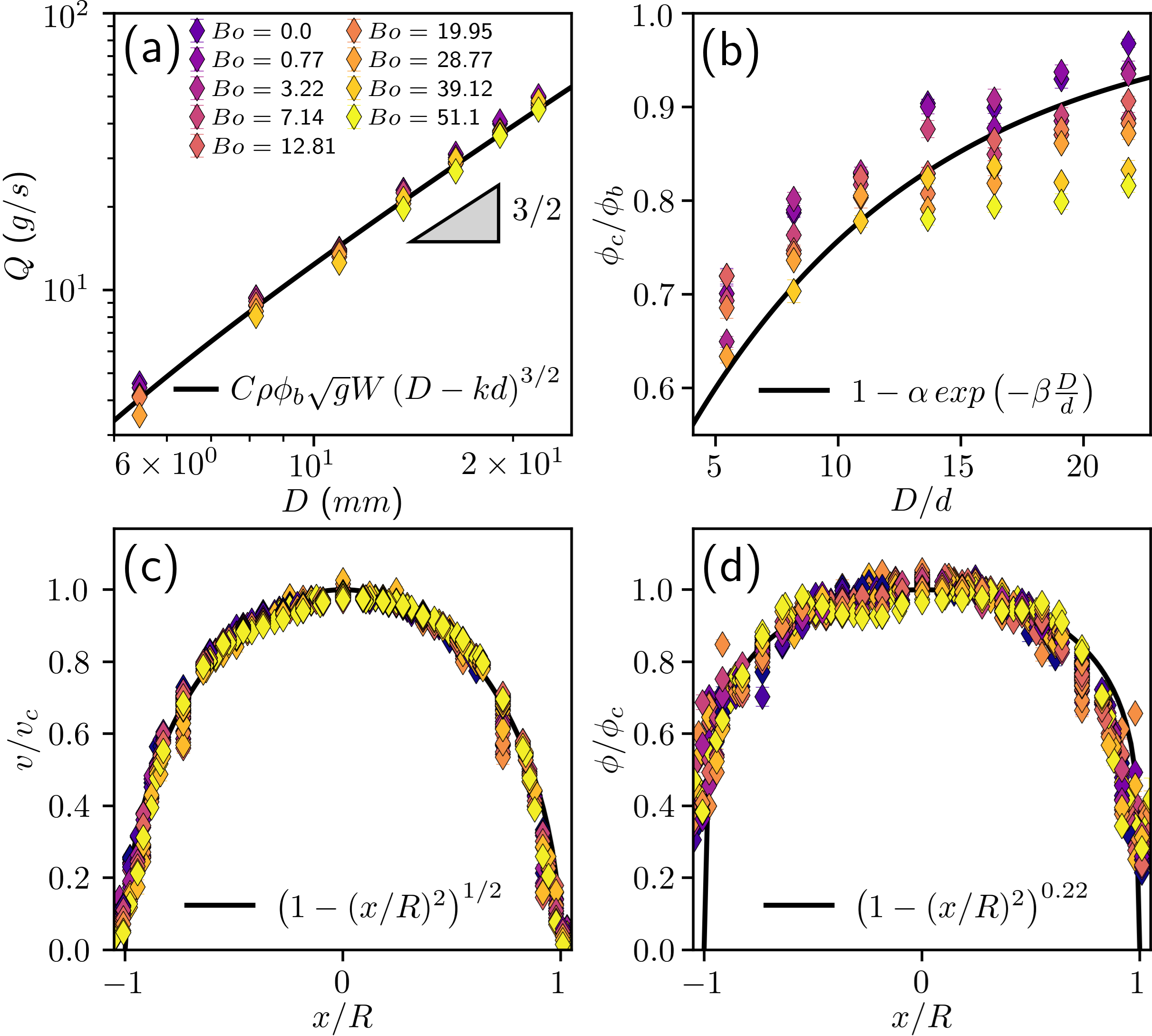}
    \vskip -0.2cm
    \caption{
    (a) Flow rates with Beverloo law for $C=0.65$, $k=1.5$ and $\phi_b=0.58$ and (b) normalized density $\phi_c(x)/\phi_b$ for all values of aperture size $D/d$ and Bond number $Bo$ ($\alpha=0.66$, $\beta=0.11$ and $\phi_b=0.58$). Normalized profiles of (c) velocity and (d) density at the outlet level normalized by their fitted center value $v_{c}$ and $\phi_c$.}
    \label{Fig:profiles}
\end{figure}

Fig.~\ref{Fig:profiles} (a,b) show the evolution of the flow rate $Q$ and the volume fraction $\phi_c$ at the center of the outlet as a function of the relative opening $D/d$. In the absence of cohesion, these quantities follow the classical laws obtained by Beverloo and Benyamine \cite{Hagen1852,beverloo1961flow, benyamine2017discharge}, respectively. $Q$ decreases slightly with $Bo$, as does $\phi_c$, which reaches an asymptotic limit at large openings, diminishing further as the Bond number increases. Cohesion has no significant effect on the velocity and compaction profiles at the outlet, which remain parabolic and self-similar \cite{janda2012flow} across all Bond numbers (see Fig.~\ref{Fig:profiles} (a)).

\begin{figure}[!h]
    \centering
    \vskip -0.2cm
    \includegraphics[width=0.48\textwidth]{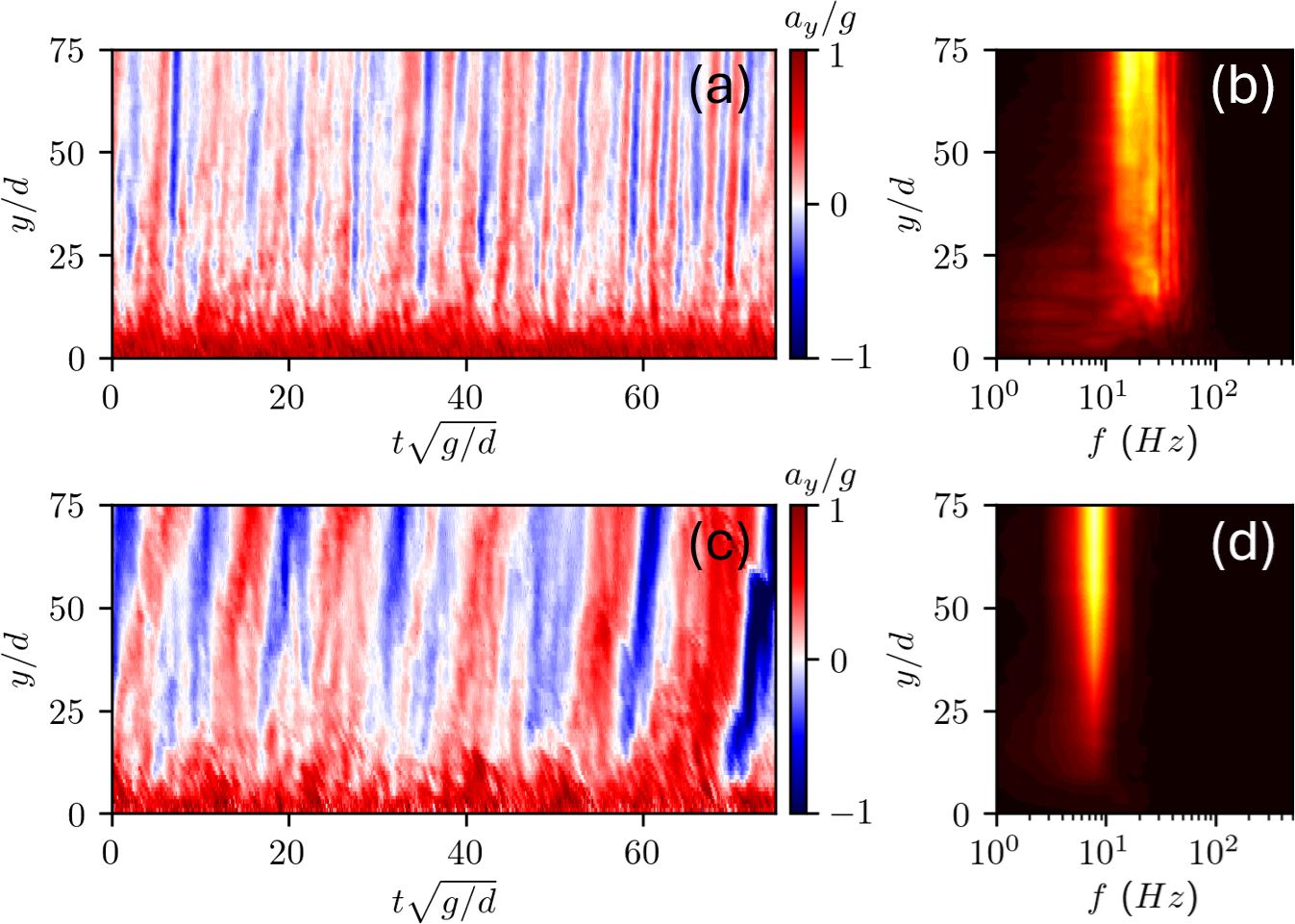}
    \vskip -0.2cm
    \caption{Spatiotemporal diagrams and corresponding power spectrum of the instantaneous acceleration $a_y(x=0)$ for $Bo\,=\,0$ (a,b)  and $Bo\,=\,51.1$ (c,d). $D\,=\,24$~mm in both cases.}
    \label{Fig:ondes2D2}
\end{figure}

The previous results are derived from the mean flow velocity during silo discharge. Here, using the high spatiotemporal resolution of our particle tracking, which provides low-noise second derivatives of particle positions, we analyze the instantaneous acceleration field to characterize flow fluctuations. Fig.~\ref{Fig:ondes2D2} (a) shows the spatio-temporal diagram of the vertical acceleration $a_y(x=0, y, t)$  of the grains located at the center of the silo during discharge in the absence of cohesion. Clear intermittency is observed, highlighting the existence of waves propagating from the opening toward the top of the silo. Fig. \ref{Fig:ondes2D2} (c) illustrates the influence of cohesion on these waves, showing that the duration of acceleration and deceleration phases increases with cohesion.

The waves are generated above the outlet, at heights ranging from $5$ to $20$~d. These acceleration waves neither block the silo nor significantly alter the mean flow rate. The flow below the region where these waves are generated exhibits particle accelerations nearly equal to the gravitational acceleration. To determine the characteristic wave frequency, we compute the power spectrum of each line of the spatio-temporal diagrams 
$a_y(x=0, y, t)$, using Welch's method (see Fig.~\ref{Fig:ondes2D2} (b,d)). The waves exhibit a clear periodicity with a well-defined frequency, about one-third of that in the non-cohesive case ($\sim 8$~Hz vs. $\sim 25$~Hz).

\begin{figure*}[!ht]
    \centering
    \includegraphics[width=1\textwidth]{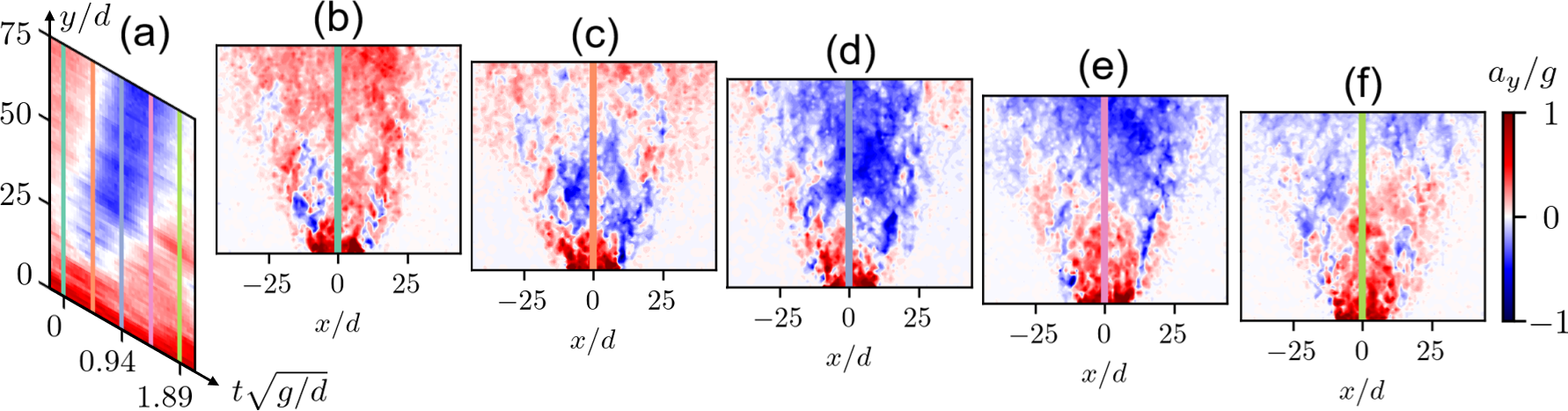}
    \caption{(a) Example of a spatiotemporal diagram with (b-f) corresponding instantaneous acceleration profiles $a_y(x,y,t)/g$ for $D\,=\,24$~mm and $Bo\,=\,0$.}
    \label{Fig:ondes2Dedge}
\end{figure*}

Examples of instantaneous snapshots (Fig.\ref{Fig:ondes2Dedge} (b-f)) show that these waves originate at the edges of the dead zones, on either side. These phenomena resemble the “silo honking” or “silo music” effect \cite{tejchman1993silo}, where sudden variations in grain flow produce waves and intermittency. Here, however, the effect does not stem from particle synchronization or wall-induced resonance, but from the periodic formation of dynamic force chains with lifetimes shorter than the flow’s characteristic timescale.

\newpage

\onecolumngrid
\twocolumngrid

\begin{figure}[!h]
    \centering
    \vskip -0.1cm
    \includegraphics[width=0.48\textwidth]{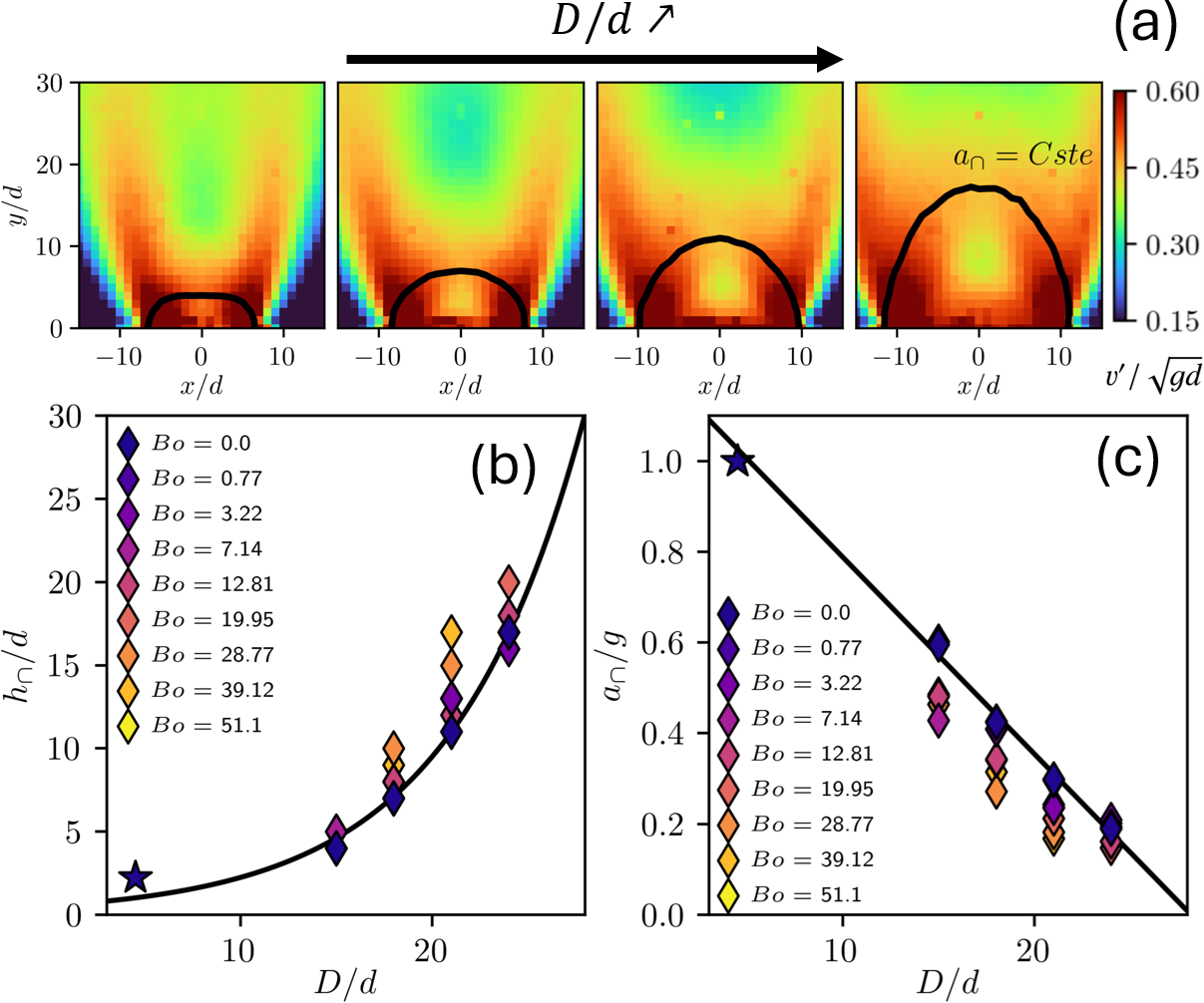}
    \vskip -0.15cm
    \caption{(a) Profiles of the normalized velocity fluctuations, $v'(x,y)/\sqrt{gd}$, for $Bo = 0$ and different aperture ratios $D/d$. (b) Dependence of the transient arch height, $h_{\cap}/d$, on the silo opening. (c) Dependence of the iso-acceleration threshold of transient arches, $a_{\cap}/g$, on the silo opening. The star symbol indicates the limiting case for which the iso-acceleration equals $g$. Solid lines in panels (b) and (c) are guides to the eye.}
    \label{Fig:fluctuations}
\end{figure}

Fig.~\ref{Fig:fluctuations} (a) presents spatial maps of the normalized velocity fluctuations $v'(x,y)/\sqrt{gd}$ within the silo near the aperture. These maps clearly reveal transient dynamic arches, which are located progressively higher in the silo as the aperture becomes larger. Remarkably, the zone of intensified fluctuations forming each arch aligns with a contour of constant acceleration (shown in black on Fig.~\ref{Fig:fluctuations} (a)), demonstrating the direct relationship between local velocity fluctuations and the acceleration field associated with these temporary structures. Fig.~\ref{Fig:fluctuations} (b,c) show, respectively, the evolution of the arch height $h_{\cap}/d$ and of the associated iso-acceleration value $a_{\cap }/g$ as a function of the aperture size. These results clearly indicate that the arches form at increasingly higher positions and become progressively less stable as the aperture increases. 

\begin{figure}[!ht]
    \centering
    \includegraphics[width=0.47\textwidth]{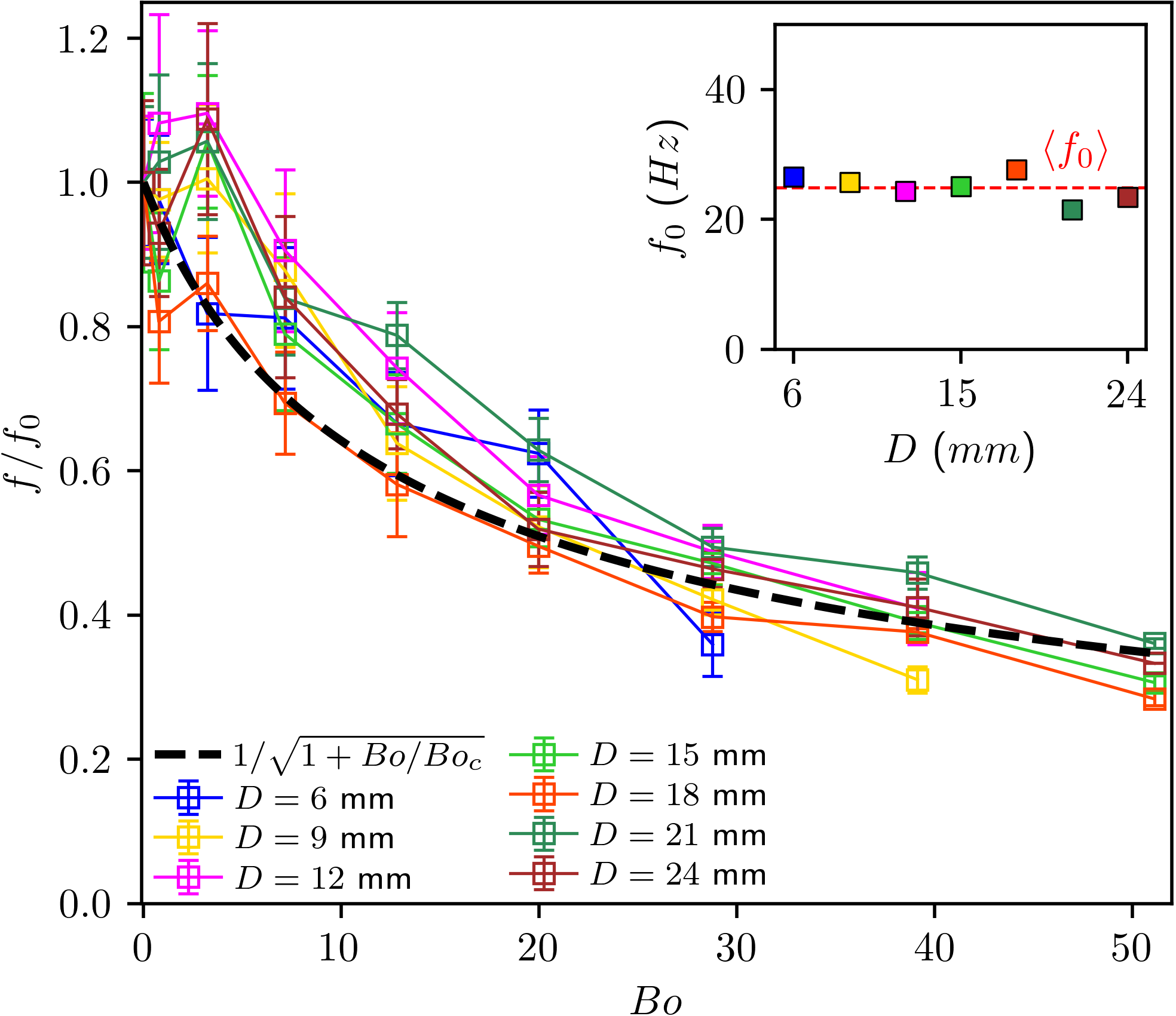}
    \vskip -0.1cm
    \caption{Evolution of the relative frequency $f/f_{0}$ of internal acceleration waves as a function of the Bond number, $Bo$, for various silo openings. Black dashed line is given by Eq. (\ref{modelcluster}) for $Bo_c=7$. Insert: $f_{0}$ as a function of aperture size, $D$. The red dashed line corresponds to the mean value $\left< f_0 \right>=25.8$~Hz.}
    \label{Fig:acceleration}
\end{figure}

These arches act as the sources of the emitted waves. We observe that even though cohesion does not have a significant influence on the flow rate, it greatly alters the emission frequency of these waves. This effect is highlighted in Fig.\ref{Fig:acceleration}, which shows the evolution of the relative frequency $f/f_0$ as a function of the Bond number, where $f_0$ is the frequency of the waves at $Bo=0$. $f/f_0$ decreases as cohesion increases, showing that deceleration waves are affected by cohesion in the bulk flow. Surprisingly, we observe that $f$ and $f_0$ do not depend on the aperture size $D$ suggesting that these waves are controlled by the grain size scale. By supposing that the magnetic field induces the formation of oriented aggregates along the vertical direction with a characteristic
length $\xi = d (1+Bo/Bo_c)$ associated with a critical Bond number $Bo_c$, the characteristic frequency related to the flow scales as $f\sim\sqrt{g/\xi}$. 
This implies that:

\begin{equation}
 f/f_{0}\sim(1+Bo/Bo_c)^{-0.5} 
 \label{modelcluster}
\end{equation}

\noindent and that the blocking frequency is determined by the free falling time under the effect of gravity of the clusters over a spatial scale equal to their diameter $\xi$. This model is shown in Fig.~\ref{Fig:acceleration} and describes the decay of the frequency relatively well. It is important to note that the model presented here is a purely geometrical model where the details of the dynamics are embedded in $f_0$, the frequency of the waves in the absence of cohesion found to be $25.8$ Hz in our experiments.

This frequency can be compared to the sound emission frequency observed in the phenomenon of booming sand dunes \cite{andreotti2004song, vriend2007solving}, which has been experimentally found to be on the order of $\alpha \sqrt{g/d}$ with $\alpha=0.4$ \cite{dagois2012singing} over a wide range of particle sizes and flow angles, giving $f_0=37.7$~Hz in our experiments. This value is very close to the one we obtained, and the difference between the two values arises from the fact that the prefactor $\alpha$ depends both on the local angle of the slope and the dissipation coefficient during collisions \cite{quartier2000dynamics} which are difficult to measure experimentally. This result suggests that the frequency of these waves in silos is associated with the velocity gradient at the surface of the dead zones, where the granular medium periodically dilates and compacts. This implies that arches form on a timescale corresponding to the time it takes for a grain to move by one diameter. The effect of cohesion can then be taken into account by assuming the formation of particle clusters whose size increases linearly with the Bond number.

Our results highlight the existence of traffic waves during silo discharge, even though the outlet flow rate remains steady. These waves are generated near the outlet by dynamic arches that locally block the flow over a characteristic timescale roughly equal to the time it takes for a grain to move by one diameter. Although this timescale is short, it scales with the grain’s free-fall time downstream of the arch, ensuring that the macroscopic flow rate remains constant. These waves propagate along the dead zones up to the top of the packing through traffic wave dynamics similar to that observed in booming sand dunes. Here, we demonstrate for the first time that the effect of cohesion can be incorporated into this mechanism by assuming the formation of clusters whose size increases with the Bond number. Our results suggest the possibility of indirectly measuring the cohesion of a granular medium by measuring the frequency of these waves. It remains an open question whether such waves occur in real cohesive powders, and how their propagation over large heights, amplification, and interaction with silo walls affect the homogeneity of silo discharge.\\

This work was financially and scientifically supported by the Lorraine Université d’Excellence (LUE) PIA project and the Institut Universitaire de France, whose contributions were essential to this investigation.
\bibliography{_biblio}

@article{mullins1974experimental,
  title={Experimental evidence for the stochastic theory of particle flow under gravity},
  author={Mullins, W.W.},
  journal={Powder Technology},
  volume={9},
  number={1},
  pages={29--37},
  year={1974},
  publisher={Elsevier}
}

@article{beverloo1961flow,
  title={The flow of granular solids through orifices},
  author={Beverloo, W.A. and Leniger, H.A. and Van de Velde, J.},
  journal={Chemical Engineering Science},
  volume={15},
  number={3-4},
  pages={260--269},
  year={1961},
  publisher={Elsevier}
}

@article{litwiniszyn1964application,
  title={An application of the random walk argument to the mechanics of granular media},
  author={Litwiniszyn, J.},
  journal = {Rheology and Soil Mechanics: Proceedings of the Grenoble Symposium, April 1--8, 1964},
  pages={82--89},
  year={1964},
  organization={Springer}
}

@article{nedderman1979kinematic,
  title={A kinematic model for the flow of granular materials},
  author={Nedderman, R.M. and T{\"u}z{\"u}n, U.},
  journal={Powder Technology},
  volume={22},
  number={2},
  pages={243--253},
  year={1979},
  publisher={Elsevier}
}

@article{medina1998velocity,
  title={Velocity field measurements in granular gravity flow in a near 2D silo},
  author={Medina, A. and Cordova, J.A. and Luna, E. and Trevino, C.},
  journal={Physics Letters A},
  volume={250},
  number={1-3},
  pages={111--116},
  year={1998},
  publisher={Elsevier}
}

@article{mankoc2007flow,
  title={The flow rate of granular materials through an orifice},
  author={Mankoc, C. and Janda, A. and Arevalo, R. and Pastor, J.M. and Zuriguel, I. and Garcimart{\'\i}n, A. and Maza, D.},
  journal={Granular Matter},
  volume={9},
  pages={407--414},
  year={2007},
  publisher={Springer}
}

@article{bazant2003theory,
  title={A theory of cooperative diffusion in dense granular flows},
  author={Bazant, M.Z.},
  journal={arXiv preprint cond-mat/0307379},
  year={2003}
}

@article{thorens2021discharge,
  title={Discharge of a 2D magnetic silo},
  author={Thorens, L. and Viallet, M. and M{\aa}l{\o}y, K. J. and Bourgoin, M. and Santucci, S.},
  journal={EPJ Web of Conferences},
  volume={249},
  pages={03017},
  year={2021},
  organization={EDP Sciences}
}

@article{benyamine2017discharge,
  title={Discharge flow of a granular media from a silo: effect of the packing fraction and of the hopper angle},
  author={Benyamine, M. and Aussillous, P. and Dalloz-Dubrujeaud, B.},
  journal={EPJ Web of Conferences},
  volume={140},
  pages={03043},
  year={2017},
  organization={EDP Sciences}
}

@article{janda2012flow,
  title={Flow rate of particles through apertures obtained from self-similar density and velocity profiles},
  author={Janda, A. and Zuriguel, I. and Maza, D.},
  journal={Physical Review letters},
  volume={108},
  number={24},
  pages={248001},
  year={2012},
  publisher={APS}
}

@article{zuriguel2019velocity,
  title={Velocity fluctuations inside two and three dimensional silos},
  author={Zuriguel, I. and Maza, D. and Janda, A. and Hidalgo, R.C. and Garcimart{\'\i}n, A.},
  journal={Granular Matter},
  volume={21},
  pages={1--9},
  year={2019},
  publisher={Springer}
}

@book{andreotti2013granular,
  title={Granular Media: Between Fluid and Solid},
  author={Andreotti, B. and Forterre, Y. and Pouliquen, O.},
  isbn={9781107034792},
  lccn={2013006089},
  url={https://books.google.fr/books?id=2ekG3NYgpqsC},
  year={2013},
  publisher={Cambridge University Press}
}

@article{cannavacciuolo2009arch,
    author = {Cannavacciuolo, A. and Barletta, D. and Dons{\`\i}, G. and Ferrari, G. and Poletto, M.},
	doi = {https://doi.org/10.1016/j.powtec.2008.10.013},
	issn = {0032-5910},
	journal = {Powder Technology},
	number = {3},
	pages = {272-279},
	title = {Arch-Free flow in aerated silo discharge of cohesive powders},
	url = {https://www.sciencedirect.com/science/article/pii/S0032591008004762},
	volume = {191},
	year = {2009}}

@article{barletta2007solid,
    author = {Barletta, D. and Donsì, G. and Ferrari, G. and Poletto, M. and Russo, P.},
    title = {Solid flow rate prediction in silo discharge of aerated cohesive powders},
    journal = {AIChE Journal},
    volume = {53},
    number = {9},
    pages = {2240-2253},
    doi = {https://doi.org/10.1002/aic.11212},
    url = {https://aiche.onlinelibrary.wiley.com/doi/abs/10.1002/aic.11212},
    year = {2007}}

@article{Hagen1852,
	Author = {Hagen, G.H.L.},
	Journal = {Bericht {\"u}ber die zur {B}ekanntmachung geeigneten {V}erhandlungen der {K}{\"o}niglich Preussischen Akademie der Wissenschaften zu Berlin},
	Pages = {35--42},
	Title = {{\"U}ber den {D}ruck und die {B}ewegung des trocknen {S}andes},
	Year = {1852}}

@article{BEN14,
  title = {Discharge flow of a bidisperse granular media from a silo},
  author = {Benyamine, M. and Djermane, M. and Dalloz-Dubrujeaud, B. and Aussillous, P.},
  journal = {Phys. Rev. E},
  volume = {90},
  issue = {3},
  pages = {032201},
  numpages = {8},
  year = {2014},
  month = {Sep},
  publisher = {American Physical Society},
  doi = {10.1103/PhysRevE.90.032201},
  url = {https://link.aps.org/doi/10.1103/PhysRevE.90.032201}
}

@article{TO01,
  title = {Jamming of Granular Flow in a Two-Dimensional Hopper},
  author = {To, K. and Lai, P.-Y. and Pak, H.K.},
  journal = {Phys. Rev. Lett.},
  volume = {86},
  issue = {1},
  pages = {71--74},
  numpages = {0},
  year = {2001},
  month = {Jan},
  publisher = {American Physical Society},
  doi = {10.1103/PhysRevLett.86.71},
  url = {https://link.aps.org/doi/10.1103/PhysRevLett.86.71}
}

@article{zuriguel2005jamming,
  title = {Jamming during the discharge of granular matter from a silo},
  author = {Zuriguel, I. and Garcimart\'{\i}n, A. and Maza, D. and Pugnaloni, L.A. and Pastor, J.M.},
  journal = {Phys. Rev. E},
  volume = {71},
  issue = {5},
  pages = {051303},
  numpages = {9},
  year = {2005},
  month = {May},
  publisher = {American Physical Society},
  doi = {10.1103/PhysRevE.71.051303},
  url = {https://link.aps.org/doi/10.1103/PhysRevE.71.051303}
}

@article{gella2017linking,
  title = {Linking bottleneck clogging with flow kinematics in granular materials: The role of silo width},
  author = {Gella, D. and Maza, D. and Zuriguel, I. and Ashour, A. and Ar\'evalo, R. and Stannarius, R.},
  journal = {Phys. Rev. Fluids},
  volume = {2},
  issue = {8},
  pages = {084304},
  numpages = {11},
  year = {2017},
  month = {Aug},
  publisher = {American Physical Society},
  doi = {10.1103/PhysRevFluids.2.084304},
  url = {https://link.aps.org/doi/10.1103/PhysRevFluids.2.084304}
}

@article{Nicolas18,
  title = {Trap Model for Clogging and Unclogging in Granular Hopper Flows},
  author = {Nicolas, A. and Garcimart\'{\i}n, A. and Zuriguel, I.},
  journal = {Phys. Rev. Lett.},
  volume = {120},
  issue = {19},
  pages = {198002},
  numpages = {5},
  year = {2018},
  month = {May},
  publisher = {American Physical Society},
  doi = {10.1103/PhysRevLett.120.198002},
  url = {https://link.aps.org/doi/10.1103/PhysRevLett.120.198002}
}

@article{zuriguel2014clogging,
  title={Clogging transition of many-particle systems flowing through bottlenecks},
  author={Zuriguel, I. and Parisi, D.R. and Hidalgo, R.C. and Lozano, C. and Janda, A. and Gago, P.A. and Peralta, J.P. and Ferrer, L.M. and Pugnaloni, L.A. and Cl{\'e}ment, E. and others},
  journal={Scientific reports},
  volume={4},
  number={1},
  pages={1--8},
  year={2014},
  publisher={Nature Publishing Group}
}

@article{xu2022modified,
  title={Modified lateral pressure formula of shallow and circular silo considering the elasticities of silo wall and storage materials},
  author={Xu, Z. and Liang, P.},
  journal={Scientific Reports},
  volume={12},
  number={1},
  pages={7069},
  year={2022},
  publisher={Nature Publishing Group UK London}
}

@article{sperl2006experiments,
  title={Experiments on corn pressure in silo cells--translation and comment of Janssen's paper from 1895},
  author={Sperl, M.},
  journal={Granular Matter},
  volume={8},
  number={2},
  pages={59--65},
  year={2006},
  publisher={Springer}
}

@article{perge2012evolution,
  title={Evolution of pressure profiles during the discharge of a silo},
  author={Perge, C. and Aguirre, M.A. and Gago, P.A. and Pugnaloni, L.A. and Le Tourneau, D. and G{\'e}minard, J.-C.},
  journal={Physical Review E},
  volume={85},
  number={2},
  pages={021303},
  year={2012},
  publisher={APS}
}

@article{tejchman1993silo,
  title={Silo-music and silo-quake experiments and a numerical Cosserat approach},
  author={Tejchman, J. and Gudehus, G.},
  journal={Powder Technology},
  volume={76},
  number={2},
  pages={201--212},
  year={1993},
  publisher={Elsevier}
}

@article{wilde2008silo,
  title={Silo music Mechanism of dynamic flow and structure interaction},
  author={Wilde, K. and Rucka, M. and Tejchman, J.},
  journal={Powder Technology},
  volume={186},
  number={2},
  pages={113--129},
  year={2008},
  publisher={Elsevier}
}

@article{carson2001silo,
  title={Silo failures: Case histories and lessons learned},
  author={Carson, J.W.},
  journal={Handbook of Powder Technology},
  volume={10},
  pages={153--166},
  year={2001}
}

@article{muite2004silo,
  title={Silo music and silo quake: granular flow-induced vibration},
  author={Muite, B.K. and Quinn, S.F. and Sundaresan, S. and Rao, K.K.},
  journal={Powder Technology},
  volume={145},
  number={3},
  pages={190--202},
  year={2004},
  publisher={Elsevier}
}

@article{andreotti2004song,
  title={The song of dunes as a wave-particle mode locking},
  author={Andreotti, B.},
  journal={Physical Review Letters},
  volume={93},
  number={23},
  pages={238001},
  year={2004},
  publisher={APS}
}

@article{vriend2007solving,
  title={Solving the mystery of booming sand dunes},
  author={Vriend, N.M. and Hunt, M.L. and Clayton, R.W. and Brennen, C.E. and Brantley, K.S. and Ruiz-Angulo, A.},
  journal={Geophysical Research Letters},
  volume={34},
  number={16},
  year={2007},
  publisher={Wiley Online Library}
}

@article{quartier2000dynamics,
  title={Dynamics of a grain on a sandpile model},
  author={Quartier, L. and Andreotti, B. and Douady, S. and Daerr, A.},
  journal={Physical Review E},
  volume={62},
  number={6},
  pages={8299},
  year={2000},
  publisher={APS}
}

@article{dagois2012singing,
  title={Singing-sand avalanches without dunes},
  author={Dagois-Bohy, S. and Courrech du Pont, S. and Douady, S.},
  journal={Geophysical research letters},
  volume={39},
  number={20},
  year={2012},
  publisher={Wiley Online Library}
}

@article{pascot2020influence,
  title={Influence of mechanical vibrations on quasi-2D silo discharge of spherical particles},
  author={Pascot, A. and Gaudel, Na. and Antonyuk, S. and Bianchin, J. and Kiesgen de Richter, S.},
  journal={Chemical Engineering Science},
  volume={224},
  pages={115749},
  year={2020},
  publisher={Elsevier}
}

@article{pascot2022discharge,
  title={Discharge of vibrated granular silo: A grain scale approach},
  author={Pascot, A. and Morel, J.-Y. and Antonyuk, S. and Jenny, M. and Cheny, Y. and Kiesgen de Richter, S.},
  journal={Powder Technology},
  volume={397},
  pages={116998},
  year={2022},
  publisher={Elsevier}
}

@article{forsyth2001effect,
  title={Effect of applied interparticle force on the static and dynamic angles of repose of spherical granular material},
  author={Forsyth, A. J. and Hutton, S. R. and Rhodes, M. J. and Osborne, C. F.},
  journal={Physical Review E},
  volume={63},
  number={3},
  pages={031302},
  year={2001},
  publisher={APS}
}

@article{vandewalle2009flow,
  title={Flow properties and heap shape of magnetic powders},
  author={Vandewalle, N and Lumay, G},
  journal={AIP Conference Proceedings},
  volume={1145},
  number={1},
  pages={135--138},
  year={2009},
  organization={American Institute of Physics}
}
\end{document}